\documentclass[preprint,12pt,authoryear]{elsarticle}

\usepackage[colorlinks=true, allcolors=blue]{hyperref}
\usepackage{xurl}

\usepackage{geometry,rotate,lscape,epsfig,booktabs,
colortbl,xspace,sectsty,amsthm,float,xy,xr,enumitem,
amssymb,amsfonts,amsbsy,amsmath,mathtools,delarray,mathtools,
 bm,color,multirow,latexsym,fancyhdr,fontenc,anysize,url,caption,subcaption,
rotating,ifthen,fancybox,fullpage,amscd,pifont,setspace,xargs,tikz-cd}
\usepackage{cleveref}

\newtheorem{prop}{Proposition}

\def\cov{{\rm cov}}

\def\Var{{\rm Var}}
\def\be{\begin{equation}}
\def\ee{\end{equation}}
\def\mT{\mathcal{T}}

\DeclareMathOperator*{\argmin}{arg\,min}

\journal{ }

\begin{document}

\begin{frontmatter}



\title{Measure Selection for Functional Linear Model} 


\author{Su I Iao and Hans-Georg M\"{u}ller\footnote{Department of Statistics, One Shields Ave., University of California, Davis, CA 95616, U.S.A.  e-mail: hgmueller@ucdavis.edu}} 


\address{Department of Statistics, University of California, Davis, One Shields Ave, Davis, 95616, CA, USA}
\begin{abstract}
Advancements in modern science have led to an increased prevalence of functional data, which are usually viewed as elements of the space of square-integrable functions $L^2$. Core methods in functional data analysis, such as functional principal component analysis, are typically grounded in the Hilbert structure of $L^2$ and rely on inner products based on integrals with respect to the Lebesgue measure over a fixed domain. A more flexible framework is proposed, where the measure can be arbitrary, allowing natural extensions to unbounded domains and prompting the question of optimal measure choice. Specifically, a novel functional linear model is introduced that incorporates a data-adaptive choice of the measure that defines the space, alongside an enhanced function principal component analysis. Selecting a good measure can improve the model's predictive performance, especially when the underlying processes are not well-represented when adopting the default Lebesgue measure. Simulations, as well as applications to COVID-19 data and the National Health and Nutrition Examination Survey data, show that the proposed approach consistently outperforms the conventional functional linear model.
\end{abstract}




\end{frontmatter}

\section{Introduction}

Functional data have become increasingly prevalent with the advancement of modern data collection technologies. Typically, functional data are considered independent and identically distributed samples representing realizations of an underlying smooth stochastic process observed at discrete time points. Over the past decades, the field of functional data analysis (FDA) has garnered significant attention, particularly in connection with the successful deployment of methods such as functional principal component analysis (FPCA) \citep{klef:73, cast:86, yao:05:1, hall:06, chen:15} and functional linear models (FLM) \citep{rams:05, yao:05:2, hall:07}. Comprehensive introductions and reviews can be found  for example in \cite{rams:05}, \cite{hsin:15}, and \cite{wang:16}. 

Both FPCA and FLM utilize the Hilbert space structure of $L^2(\mathcal{T})$, which is conventionally equipped with the Lebesgue measure to facilitate computation, where $\mathcal{T}$ denotes the continuum of interest for the functional data. FLM is often implemented using an FPCA-based approach \citep{yao:05:2, hall:06, nadi:13, imai:18}, where FPCA is first applied to decompose functional predictors into orthogonal principal components. These principal components serve as low-dimensional representations that are subsequently used as covariates in a regression model. However, while this approach is widely used in applications \citep{lian:15, chen:24, iao:24, zhou:24}, it may not always provide the most effective representation of functional data. The success of FPCA-based FLM largely depends on the efficient representation of the coefficient function of FLM by the leading functional principal components \citep{cai:12}. In practice, the principal components of $X$ may not align well with the structure of the coefficient function, leading to suboptimal predictive performance. This issue parallels limitations observed in principal component regression \citep{joll:82} and singular value decomposition techniques for linear inverse problems \citep{dono:95}. Notably, when low-variance components of $X$ carry non-negligible predictive power, discarding them can degrade model performance. These considerations motivate the development of alternative eigensystem constructions aimed at improving predictive accuracy and model interpretability.

A promising approach to address this issue is to introduce weighting schemes in functional data analysis. Prior research has explored weighted methods in FPCA \citep{leng:04, tals:20}, as well as in clustering and classification \citep{chen:14, roma:20}. By defining inner products with respect to an alternative measure, the resulting eigensystem can potentially yield a more effective representation of the coefficient function. Despite these developments, the application of weighted methodologies to functional linear models remains an open research question.

In this work, we propose a novel weighted functional linear model (wFLM) with functional predictors and scalar responses based on a data-driven measure. This framework is designed to operate in a Hilbert space equipped with a general measure, transcending beyond the classical default Lebesgue measure, and applies to both bounded and unbounded domains. By incorporating a general measure, the proposed approach enables a more flexible representation of functional data, leading to improved model interpretability and predictive performance.  In addition to optimizing eigensystem alignment, the weighting approach that we propose here conveys additional benefits when dealing with infinite domains. When the domain is unbounded $\mT = [0,\infty)$, the space $L^2(\mathcal{T})$ imposes major constraints, as commonly used functions, like polynomials, are not situated in this space, while they are square-integrable when $\mathcal{T}$ is finite. If trajectories  $X$ do not lie in $L^2(\mathcal{T})$, traditional functional data analysis techniques such as FPCA and FLM are not applicable. Adopting a weighting scheme might also reflect that not all regions of a function's domain are equally important or relevant for the analysis. Changing the uniform reference measure may be interpreted as emphasizing or downplaying the variability at some subdomains of the stochastic processes.


The rest of this paper is organized as follows. In Sections 2, we introduce the weighted functional principal component analysis and weighted functional linear model. The data-adaptive measure selections are established in Section 3. Simulations are shown in Section 4. Applications for COVID-19 data and the National Health and Nutrition Examination Survey data are discussed in Section 5.

\section{Methodology}

In this section, we introduce weighted functional principal component analysis (wFPCA) and then proceed to the wFPCA-based weighted functional linear model (wFLM).

\subsection{Weighted functional principal component analysis}
Revisiting classical functional principal component analysis (FPCA), let  $X(t)$ be a square integrable stochastic process on $\mT$ for which one has $n$ independent copies $X_i(t)$, $i = 1,\ldots,n$. Mean and covariance function of $X$ are 
\begin{equation}\label{eq:X_mean_cov}
	\mu_X(t) = E\{X(t)\} \text{ and } C_{XX}(s,t)=E[ \{X(s) - \mu_X(s)\}\{X(t) - \mu_X(t)\}].
\end{equation}
By Mercer's Theorem (see Theorem 4.6.5 in \citet{hsin:15}), the spectral decomposition of the covariance function $C_{XX}(s,t)$ is 
$$C_{XX}(s,t)=\sum_{k=1}^\infty\rho_k\phi_k(s)\phi_k(t),\quad s,t\in\mathcal{T},$$
where $\rho_1>\rho_2>\cdots>0$ are the eigenvalues and $\{\phi_k\}_{k=1}^\infty$ are the corresponding eigenfunctions of the auto-covariance operator.  The latter form an orthonormal system on $L^2(\mT)$ with respect to inner products based on the Lebesgue measure. The Karhunen-Lo\`eve representation  implies that the $i$th random curve can be represented as
$$
X_i(t)=\mu_X(t)+\sum_{k=1}^{\infty} \xi_{i k} \phi_k(t), \quad t \in \mT,
$$
where the principal component scores $\xi_{i k}=\int_{\mT}\left\{X_i(t)-\mu_X(t)\right\} \phi_k(t) dt$ are uncorrelated random variables with zero mean and variances $E(\xi_{i k}^2)=\rho_k$.

An extension of classical FPCA involves defining  inner products with respect to more general measures, 
\begin{equation*}
	\langle f, g\rangle =\int_{\mathcal{T}} f(t) g(t) d \nu(t),
\end{equation*}
where $\nu$ is an absolutely continuous measure with respect to  Lebesgue measure and $f, g$ are $\nu$-square integrable functions on the domain $\mT$, see, e.g.,  \citet{leng:04, chen:14, tals:20}. 
 The Radon-Nikodym theorem \citep{ash:00} ensures the existence of a measurable function $w(t) = d\nu(t)/dt$, $t\in\mT$, which we refer to as weight function. The weight function $w$ is assumed to reside in the space 
\begin{equation}\label{eq:Wset}
	\mathcal{W} = \left\{w: \mT \mapsto [0,\infty);  \int_{\mathcal{T}}w(t)dt = \int_{\mathcal{T}}d\nu(t) = 1\right\}.
\end{equation}

To define the weighted FPCA, we assume that $X(t)$ is square integrable on $\mT$ with respect to $\nu$, i.e., $\int_\mT X^2(t) d\nu(t)<\infty$ and introduce a new square integrable process with respect to Lebesgue measure, 
$$Z = \sqrt{w} (X - \mu_X),$$ 
where $Z$ has $E(Z(t)) = 0$ and covariance function  $C_{ZZ}(s,t) = E\{Z(s)Z(t)\}$. The spectral decomposition of the covariance $C_{ZZ}$ with respect to Lebesgue measure is
\begin{equation}\label{eq:covZZ}
	C_{ZZ}(s,t)=\sum_{k=1}^\infty\rho_{wk}\phi_{Zk}(s)\phi_{Zk}(t),\quad s,t\in\mathcal{T},
\end{equation}
with eigenvalues  $\rho_{w1} > \rho_{w2} >\cdots>0$  and eigenfunctions $\{\phi_{Zk}\}_{k=1}^\infty$.  
The Karhunen-Lo\`eve representation of the process $Z$ is
\begin{align*}
	Z(t) =& \sum_{k=1}^\infty \xi_{wk}\phi_{Zk}(t), \label{eq:Z_KL}\\
	\xi_{wk} =& \int_{\mT} Z(t) \phi_{Zk}(t)dt = \int_{\mT} \{X(t)-\mu_X(t)\} \phi_{wk}(t) d\nu(t),
\end{align*}
where $\xi_{wk}$ is the $k$th principal component score of $Z$ and $\phi_{wk}(t) = \phi_{Zk}(t) / \sqrt{w(t)}$. To ensure $\phi_{wk}(t)$ is well-defined, we set $1/\sqrt{w(t)} = 0$ if $w(t) = 0$. This leads to the following proposition, which appeared previously in \citet{leng:04}.

\begin{prop}\label{prop:eigen-system-Z}
	Given a probability measure $\nu$ that is absolutely continuous with respect to the Lebesgue measure, where $d\nu(t) = w(t)dt$, 
	and a stochastic process $X(t)\in L^2(\mT, \nu)$, the process $Z = \sqrt{w}(X - \mu_X)$ is mean zero and square integrable with respect to the Lebesgue measure. For the eigenvalues and eigenfunctions $\{\rho_{wk}, \phi_{Zk}\}_{k=1}^\infty$ of the process $Z$ with respect to Lebesgue measure as per \eqref{eq:covZZ} and the functions 
	$$\phi_{wk} = \frac{\phi_{Zk}}{\sqrt{w}}, \,\, k=1,2,\dots,  \text{  where } \frac{1}{\sqrt{w}} = 0 \text{ if } w = 0,$$ it holds that the $\{\rho_{wk},\phi_{wk}\}_{k=1}^\infty$ form the eigensystem of the original process $X$ with respect to the probability measure $\nu$. The Karhunen-Lo\`eve representation of $Z$ is given by
    \begin{equation}
        Z(t) = \sum_{k=1}^\infty \xi_{wk}\phi_{Zk}(t),  \label{eq:Z_KL}
    \end{equation}
    with principal component scores 
    \begin{equation}
	\xi_{wk} = \int_{\mT} Z(t) \phi_{Zk}(t)dt = \int_{\mT} \{X(t)-\mu_X(t)\} \phi_{wk}(t) d\nu(t),\,\, k=1,2,\dots\,.\label{eq:xi_wk}
    \end{equation}
    The scores $\xi_{wk}$ can equivalently be interpreted as principal component scores of the process $Z$ under the Lebesgue measure or as principal component scores of the process $X$ under the probability measure $\nu$.
    
    
\end{prop}

All proofs are provided in the Supplementary Material. Given a general measure $\nu$, Proposition \ref{prop:eigen-system-Z} yields an easily implementable approach to obtain the Karhunen-Lo\`{e}ve expansion of processes $X$ and the weighted FPCA of a process $X$ in $L^2(\mathcal{T}, \nu)$. With the measure $\nu$ and random samples $\{X_{i}\}_{i=1}^n$, one can follow the estimation procedures outlined in \citet{yao:05:1} and \citet{zhan:16} to obtain estimates $\hat{\mu}_X$, $\hat{C}_{ZZ}$ and further derive estimates $\hat{\rho}_{wk}$, $\hat{\phi}_{Zk}$,  $\hat{\xi}_{wk}$ and  $\hat{\phi}_{wk}$ for the corresponding targets indexed by $k = 1, \ldots, M$ (where $M$ is the number of included eigenfunctions, which can be chosen by leave-one-out cross-validation, see Section~\ref{sec:selection}). 

Proposition~\ref{prop:eigen-system-Z} relies on two key assumptions that are standard and well-motivated in functional data analysis: (1) The stochastic process $X(t)$ resides in the Hilbert space $L^2(\mathcal{T}, \nu)$, i.e., it is square integrable with respect to the general measure $\nu$. This assumption ensures the existence of well-defined mean and covariance functions and guarantees the applicability of the Karhunen-Lo\`eve expansion and the Hilbert space structure of 
$L^2(\mathcal{T}, \nu)$ provides the basis for eigen-analysis, including completeness, a well-defined inner product, and the existence of an orthonormal basis. (2)  The measure $\nu$ is absolutely continuous with respect to the Lebesgue measure, so that the Radon-Nikodym derivative $w(t) = d\nu(t)/dt$ exists. This is a mild condition, commonly satisfied in practical applications where the weighting function is derived from data or design considerations. In particular, it is satisfied by our proposed data-driven measure, which is constructed to be absolutely continuous by design; see Section~\ref{sec:selection} for further details. Intuitively, absolute continuity ensures that $\nu$ does not assign positive mass to any set that has zero Lebesgue measure, so no information carried by $\nu$ is lost when working with Lebesgue integrals, allowing for the analysis of the transformed process $Z = \sqrt{w}(X - \mu_X)$  in the standard $L^2(\mathcal{T})$ setting.

More generally, if we consider two  measures $\nu_1$ and $\nu_2$,  where  $\nu_2$ is absolutely continuous with respect to $\nu_1$, with Radon-Nikodym  derivative $d\nu_2/d\nu_1$,  one can conduct a weighted FPCA within the space $L^2(\mT, \nu_2)$ by means of the space $L^2(\mT, \nu_1)$. The following proposition extends Proposition \ref{prop:eigen-system-Z} to this more general setting.

\begin{prop}\label{prop:eigensystem}
	Given two general measures $\nu_1$, $\nu_2$ and a stochastic process $X(t)\in L^2(\mT, \nu_2)$, if we assume $\nu_2$ is absolutely continuous with respect to $\nu_1$, then $Z = \sqrt{\frac{d\nu_2}{d\nu_1}}(X - \mu_X)$ belongs to $L^2(\mT, \nu_1)$. Denote the eigenvalues and eigenfunctions of the process $Z$ with respect to $\nu_1$ as $\{\lambda_{wk}, \psi_{Zk}\}_{k=1}^\infty$ and define a new function
	$$\psi_{wk} = \psi_{Zk}/\sqrt{\frac{d\nu_2}{d\nu_1}}.$$
Then, the eigensystem of $X$ within $L^2(\mT, \nu_2)$ is
$\{\lambda_{wk},\psi_{wk}\}_{k=1}^\infty,$
and the $k$th principal component score of the process $X$ with respect to the  measure $\nu_2$ is
$$\zeta_{wk} = \int_\mT Z(t) \psi_{Zk} (t) d\nu_1(t) =  \int_\mT \{X(t) - \mu_X(t)\} \psi_{wk} (t) d\nu_2(t).$$
\end{prop}

Proposition~\ref{prop:eigensystem} generalizes  Proposition~\ref{prop:eigen-system-Z} by establishing a mapping between eigensystems defined under two arbitrary measures, provided one of these measures absolutely continuous with respect to the other. This result enables the construction of a weighted FPCA framework in $L^2(\mathcal{T}, \nu_2)$ by leveraging the eigendecomposition of a rescaled process $Z$ in a potentially simpler or more tractable space $L^2(\mathcal{T}, \nu_1)$. The key idea is that the Radon-Nikodym derivative \( d\nu_2/d\nu_1 \) determines how the geometry of the space, and hence the structure of the principal components, transforms across different weighting schemes.

When in Proposition~\ref{prop:eigensystem} $\nu_1$ is the  Lebesgue measure and $\nu_2 = \nu$ is an absolutely continuous measure with Radon-Nikodym derivative $w(t) = d\nu(t)/dt$,  Proposition~\ref{prop:eigen-system-Z} emerges as a special case, where 
 the transformed process $Z(t) = \sqrt{w(t)}\{X(t) - \mu_X(t)\}$ and the reweighted eigenfunctions $\phi_{wk}(t) = \phi_{Zk}(t)/\sqrt{w(t)}$ match those in Proposition~\ref{prop:eigen-system-Z}.



\subsection{Weighted functional linear model}
Consider a general measure $\nu$ which is absolutely continuous with respect to the Lebesgue measure, such that there exists a weight function $w(t) = d\nu(t)/dt \in \mathcal{W}$. Let $(X, Y)$ be a random pair in $L^2(\mT, \nu)\times \mathbb{R}$, where $X$ is a functional predictor and $Y$ a scalar response, where $\mu_Y = E(Y)$ and $\mu_X=E(X)$, and variance 
$\sigma_Y^2 = \Var(Y)$ and covariance $C_{XX}$  as per \eqref{eq:X_mean_cov}. Suppose $\{X_i, Y_i\}_{i=1}^n$ are $n$ independent realizations of $(X,Y)$. In this section, we consider a weighted functional linear model (wFLM) in which $(X,Y)$ are generated by the model
\begin{equation*}
    E[Y | X] = \beta_0 + \int_\mT X(t) \beta(t) d\nu(t).
\end{equation*}
Here the regression function $\beta(t)$ is smooth and square integrable, i.e.,  $\int_\mathcal{T}\beta^2(t)d\nu(t) < \infty$. 

Centering predictor processes $X$, the functional linear regression model becomes
\begin{equation*}\label{true centered}
    E[Y | X] = \mu_Y + \int_\mathcal{T} \{X(t)-\mu_X(t)\} \beta(t) d\nu(t).
\end{equation*}
Consider the transformed processes $Z=\sqrt{w}(X-\mu_X)$, the wFLM is equivalent to
\begin{equation}\label{eq:EYZ}
    E[Y | X] = \mu_Y +\int_\mathcal{T} Z(t) \beta_w(t)dt, 
\end{equation}
where $\beta_w(t) = \beta(t)\sqrt{w(t)}$. The regression parameter function $\beta_w$ can be represented as \citep{yao:05:2, hall:07} 
\begin{equation}\label{eq:beta_w}
    \beta_w(t) = \sum_{k=1}^{\infty} \frac{E\{\xi_{wk} (Y - \mu_Y)\}}{E(\xi_{wk}^2)} \phi_{Zk}(t) = \sum_{k=1}^{\infty}\rho_{wk}^{-1}\sigma_{kY} \phi_{Zk}(t) = \sum_{k=1}^{\infty} \beta_k \phi_{Zk}(t),
\end{equation}
where $\{\rho_{wk}, \phi_{Zk}\}_{k=1}^{\infty}$ is the  eigensystem of the process $Z$ as per \eqref{eq:covZZ}, $\xi_{wk}$ are the $k$th principal component scores of the process $Z$ as per \eqref{eq:xi_wk}, $\sigma_{kY} = E\{\xi_{wk} (Y - \mu_Y)\}$ and $\beta_k = \rho_{wk}^{-1}\sigma_{kY}$. Transforming $\beta_w$ back to $\beta$, we obtain the representation 
\begin{equation}
        \beta(t) = \sum_{k=1}^{\infty} \beta_k \phi_{wk}(t).
\end{equation}

One can use a well-established local linear smoothing approach to obtain an estimate $\hat{C}_{YZ}(t)$ of the cross-covariance surface 
\begin{equation*}
	C_{YZ}(t) = \cov(Y, Z(t)) = \sum_{k=1}^\infty E\{\xi_{wk}(Y-\mu_Y)\}\phi_{Zk}(t).
\end{equation*}
This leads to the estimators
\begin{equation}
	\hat\beta_w(t) = \sum_{k = 1}^M \hat\beta_k\hat\phi_{Zk}(t)\quad \text{and}\quad \hat\beta(t) = \sum_{k = 1}^M \hat\beta_k\hat\phi_{wk}(t),
\end{equation}
where $M$ is number of included eigen-components, which is a  tuning parameter, $\hat{\beta}_k = \hat\rho_{wj}^{-1}\hat{\sigma}_{kY}$, $\hat{\sigma}_{kY} = \int_{\mT}\hat{C}_{YZ}(t)\hat\phi_{Zk}(t)dt$. Further details about this smoothing approach to obtain estimates of the eigen-components and coefficient functions can be found, e.g., in \citet{yao:05:2}.

To predict the scalar response $Y^*$  from a new predictor trajectory $X^*$, we ultilize the equation \eqref{eq:EYZ}, the basis representation of $\beta_w(t)$ as per \eqref{eq:beta_w} and the orthonormality of the $\{\phi_{Zk}\}_{k\geq 1}$. The prediction of the response can be obtained via the conditional expectation
\begin{equation}
	E[Y^* | X^*] = \mu_Y + \sum_{k=1}^\infty \rho_{wk}^{-1}\sigma_{kY} \xi^*_{wk},
\end{equation}
where 
$$\xi^*_{wk} = \int_{\mathcal{T}}(X^*(t) - \mu_X(t))\phi_{wk}(t) d\nu(t) = \int_{\mT}Z^*(t)\phi_{Zk}(t)dt$$
is the $j$th functional principal component score of the predictor trajectory $X^*$. The quantities $\mu_Y$, $\mu_X$, $\rho_{wk}$, $\sigma_{kY}$ can be estimated from the data, as described in \citet{yao:05:1, yao:05:2} and \citet{zhan:16}.  

\section{Choosing the weight function for the functional linear model}\label{sec:selection}

So  far the weight function, $w$ was assumed to be given.  In practical applications, selecting a good  weight function from the available data is crucial. Ideally, we aim to find the optimal weight function within  a  set $W$ of potential weight functions  as per \eqref{eq:Wset}. The objective is to minimize the cross-validation error, 
\begin{equation}\label{CVE}
	w^* = \argmin_{w\in W} \text{CVE}(w) = \argmin_{w\in W} \frac{1}{n}\sum_{i=1}^n\left(Y_i - \hat{Y}_{i,M}^{(-i)}\right)^2,
\end{equation}
where $\hat{Y}^{(-i)}_{i,M} = \hat\mu_Y^{(-i)} + \sum_{k=1}^M\hat\beta_k^{(-i)}\hat{\xi}_{i,Zk}^{(-i)}$ is the cross-validation prediction for the $i$th subject and $\hat\beta_k^{(-i)} = \hat\sigma_{kY}^{(-i)}/\hat{\rho}_{wk}^{(-i)}$ is the estimate of $\beta_k$ as per \eqref{eq:beta_w}. Here, the superscript $(-i)$ denotes leave-one-out estimation, where the $i$th sample is omitted from the estimation process. To accomplish this goal, we present two practical approaches for selecting optimal weight functions tailored to different types of domain $\mathcal{T}$, aiming to  up-weigh or down-weigh subdomains that are more or less important for obtaining good predictions when applying the functional linear model.

\subsection{Step function approach on the finite domain $\mathcal{T} = [0,1]$}

Finding an analytical solution for the optimal weight function in Equation \eqref{CVE} is challenging. To efficiently obtain approximate solutions for \eqref{CVE} in practical applications, we employ a dyadic splitting algorithm \citep{leng:04}.  For the sake of completeness, details about this algorithm are included in the Supplementary Material. This algorithm results in  weight functions that are step functions. 

We search for the optimal weight function within the subset $W_{\text{step}} \subset W$, where 
\begin{equation}
	W_{\text{step}} = \{w(t)=\sum_{l=1}^{2^K} c_l\cdot 1_{t\in [\frac{l-1}{2^K},\frac{l}{2^K})}: \,\, c_l\in \mathbb{R},\,\, \int_{\mathcal{T}} w(t)dt=1 \}.
\end{equation}
Here,$2^K$ is the number of steps and $K$ is the number of times that we split the interval. To ensure that the resulting weight function is interpretable and has no abrupt jumps, we  consider a penalized cross-validation score \eqref{PCVS}, 
\begin{equation}\label{PCVS}
	w^* = \argmin_{w\in W_{\text{step}}} \text{PCVS}(w) = \argmin_{w\in W_{\text{step}}} \frac{\sum_{i=1}^n\left(Y_i - \hat{Y}_{i,M}^{(-i)}\right)^2}{\sum_{i=1}^n\left(Y_i - \bar{Y}\right)^2} + \lambda_1 TV(w) + \lambda_2 \frac{\int_{\mathcal{T}} I\{w(t)\ne 0\}dt}{|\mathcal{T}|},
\end{equation}
where $TV(w) = \sum_{l}^{2^K-1} |c_{l+1} - c_{l}|$ is the total variation of $w$ and $M$, $\lambda_1$, $\lambda_2$ are tuning parameters, $M$ denoting the number of included components. 

For the selection of tuning parameters, we employed cross-validation to simultaneously select $M, \lambda_1$ and $\lambda_2$. 
To ensure computational efficiency, we limit the number of candidate values for $M$, $\lambda_1$, and $\lambda_2$ to expedite the cross-validation procedure.  For $M$, we consider candidate values  ranging from 1 to $\text{M}_{Leb}$, where $\text{M}_{Leb}$ is the best value for $M$ in the FLM under the Lebesgue measure according to the cross-validation error;  for  $\lambda_1$ and $\lambda_2$, we consider  the values 0, 0.5 and 1. For a comprehensive sensitivity analysis for the choice of the tuning parameters $\lambda_1$ and $\lambda_2$ we  refer to Section S.8 of the Supplementary Material.


\subsection{Parametric density approach on the infinite domain $\mathcal{T} = [0,\infty)$}


For  infinite domains, we adopt a parametric approach for selecting the weight function $w$.  Specifically, we consider density functions whose support aligns with $\mathcal{T} = [0,\infty)$ or $\mathcal{T} = (-\infty,\infty)$ to ensure that the resulting weighted $L^2$ space remains well-defined. We focus on  the case $\mathcal{T} = [0,\infty)$, as extensions to the case $\mathcal{T} = (-\infty,\infty)$ are analogous. For  $\mathcal{T} = [0,\infty)$, suitable choices include distributions from the exponential family, such as the exponential, half-normal, gamma and truncated normal distributions. These parametric choices incorporate prior knowledge or a desired emphasis on  specific subregions of the domain. In our applications, we focus on the exponential density  $w(t; \lambda) = \lambda e^{-\lambda t}$ for $t \in [0,\infty)$ due to its interpretability, single parameter $\lambda>0$ which controls the decay rate and its strong empirical performance.  
The exponential density places more weight near the origin and decays monotonically, which is often appropriate in functional data where signal strength may diminish over time. We also considered the half-normal distribution in our simulation studies, where $w(t; \sigma) = \sqrt{\frac{2}{\pi \sigma^2}} \exp\left(-\frac{t^2}{2\sigma^2}\right)$ for $t \in [0,\infty)$. The half-normal distribution also defines decreasing weights over $[0, \infty)$, and while its rate of decay differs from that of the exponential distribution, both densities asymptotically approach zero as $t \to \infty$. As demonstrated in Section~\ref{subsec:unbounded}, the predictive performances of exponential and half-normal weights are comparable, suggesting robustness to specific choices. For each choice we selected the optimal parameter via cross-validation using the criterion in Equation~\eqref{CVE}.

While the exponential density is emphasized in our applications and the half-normal is included in our simulations, the proposed framework is not restricted to these choices. Weight functions derived from other parametric distributions such as the gamma distribution could also be incorporated. These alternatives offer additional flexibility. Both gamma and truncated normal distributions may place more emphasis on mid- or late-domain regions rather than near the origin, which may be beneficial in settings where important information is concentrated away from $t = 0$. Although these densities differ in shape near the center, they all exhibit exponential decay as $t \to \infty$, ensuring stability over unbounded domains. Thus, the gamma distribution and truncated normal distribution may be suitable alternatives when emphasizing mid-to-late domain regions is desirable. Among these different options, the exponential density provides a computationally efficient and conceptually straightforward baseline. Nonetheless, the proposed framework is flexible and can accommodate weight functions derived from other parametrically specified distributions.

\section{Simulation studies}

\subsection{Simulations on $\mathcal{T} = [0,1]$}\label{subsec:bounded}
We conducted simulation studies evaluating weight functions for two distinct measures: the Lebesgue measure (uniform density) and an optimal measure as approximated by a step function. These investigations comprised two separate simulation scenarios, each encompassing $Q=1000$ Monte Carlo runs. In each scenario, we considered $12$ settings with $n$ ranging from ${50, 100, 200, 500}$ i.i.d. pairs, consisting of a response scalar and a predictor trajectory, as well as varying numbers of measurements per predictor trajectory  $N=20,50,100$; the locations $t_{ij}$ where these measurements were taken were equidistant within the interval $[0,1]$.

The predictor trajectories, denoted as $X_i(\cdot)$ with corresponding noisy measurements $X_{ij}$, were generated as follows. For both scenarios, the simulated processes $X$ had mean function  $\mu_X(t)=2t-5\cos (2\pi t)$ and covariance functions were constructed using 10 eigenfunctions $\psi_k(t)$ such that 
$$\psi_k(t) = \begin{cases}
	\sqrt{2}\cos\left(k\pi t\right), & k \in  \{1,3,5,7,9\}, \\
      \sqrt{2}\sin\left((k-1)\pi t\right), & k \in  \{2,4,6,8,10\}.
\end{cases}$$

For Scenario 1 we chose eigenvalues  $\rho_k=10\times 0.5^{10-k}$ for $k=1,\ldots,10$. 
We generated functional principal component scores $\xi_{ik}$ from $\mathcal{N}\left(0, \rho_k\right)$ and obtained the predictor measurements 
\be \label{fmsim} X_{ij}=\mu_{X}(t_{ij})+\sum_{k=1}^{10}\xi_{ik}\psi_k(t_{ij})+\epsilon_{ij},
\ee 
where the additional measurement errors $\epsilon_{ij}$ followed a normal distribution with mean 0 and variance $0.5^2$, $j = 1,\ldots, N$ and $i = 1,\ldots, n$. The scalar responses were generated according to  $Y_i = \int_{0}^{1}\beta(t)X_i(t)dt + e_i = \sum_{k=1}^{10}\beta_k\xi_{ik}+e_i,$ where $\beta(t)=\sum_{k=1}^{10}\beta_k\psi_k(t)$, with $\beta_k=5\times 0.5^{k-1}$ for $k=1,\ldots,10$, and the additional measurement errors for the responses $e_i$ followed a normal distribution with mean 0 and variance $0.5^2$. Processes $X$ and all errors were independent in both Scenario 1 and Scenario 2. 

In Scenario 2 we chose eigenvalues  $\rho_k=10\times 0.5^{k-1}$ for $k=1,\ldots,10$. 
We generated FPC scores $\xi_{ik}$ from $\mathcal{N}\left(0, \rho_k\right)$, and calculated the predictor measurements $X_{ij}$
again as in \eqref{fmsim}, where all errors were obtained in the same way as in Scenario 1.  
Here the scalar responses were generated as  $Y_i = \int_{0}^{1}\beta(t)X_i(t)w(t)dt + e_i$, where 
$\beta(t)=2 + 3t - 3\sin(\pi t)$, with $e_i$ as in Scenario 1 and the weight function $w(t)$ was specified as
\begin{equation*}
	 w(t) =  \begin{cases}
	 	0, & t \in  (-\infty,1/4)\cup (1,+\infty);\\
      1/6, & t\in [1/4, 1/2);\\
      1/3, & t \in [1/2, 3/4);\\
      1/2, & t\in [3/4, 1].\\
	 \end{cases}
\end{equation*}

For the $q$th Monte Carlo run, we generated 100 new noisy predictors  $X_{ij,q}^*$ and 100 corresponding noise-free responses  $Y_{i,q}^*$. We evaluated the predictive performance using the average mean squared prediction error (AMSPE)
\begin{equation}\label{eq:MSPE}
	\text{AMSPE} = \frac{1}{Q}\sum_{q=1}^{Q}\frac{1}{100}\sum_{i=1}^{100}(\hat{Y}^{*}_{i,q} - Y_{i,q}^*)^2,
\end{equation}
where $\hat{Y}_{i,q}^*$ represented the predicted responses estimated by either FLM or wFLM.

Table \ref{tab:bounded} presents the results for both scenarios. In Scenario 1, wFLM (step) consistently outperformed FLM (Lebesgue) in terms of AMSPE, with larger gains observed for increased sample sizes and measurement points. These results suggest that when the coefficient function $\beta(\cdot)$ cannot be efficiently represented using the leading functional principal components, using the default Lebesgue measure in the FLM may be suboptimal and a more general step function-approximated measure may entail a more suitable eigensystem to efficiently represent the regression parameter function $\beta$, especially when sample size $n$ and number of measurement points $N$ are relatively large. Furthermore, in Scenario 2, wFLM (step) also demonstrated superior predictive performance across all settings, with notable improvements when $n = 500$, achieving reductions in MSPE of $51\%$, $78\%$ and $82\%$ for numbers of measurements $N=20, 50, 100$.


\begin{table}[tb]
\centering
	\begin{tabular}{cccccc}
\toprule
Scenario & Method & $n$ & $N=20$ & $N=50$ & $N=100$\\
\midrule
\multirow{8}{*}{I} & FLM & \multirow{2}{*}{50} & 1.004 (0.215) & 1.000 (0.214) & 1.001 (0.217)\\
& wFLM &  & 0.905 (0.218) & 0.813 (0.197) & 0.815 (0.200)\\
\cmidrule{2-6}
& FLM & \multirow{2}{*}{100} & 0.975 (0.142) & 0.973 (0.142) & 0.973 (0.142)\\
& wFLM &  & 0.846 (0.154) & 0.722 (0.134) & 0.703 (0.137)\\
\cmidrule{2-6}
& FLM & \multirow{2}{*}{200} & 0.967 (0.100) & 0.967 (0.101) & 0.968 (0.100)\\
& wFLM &  & 0.821 (0.118) & 0.668 (0.088) & 0.644 (0.082)\\
\cmidrule{2-6}
& FLM & \multirow{2}{*}{500} & 0.958 (0.060) & 0.958 (0.061) & 0.958 (0.060)\\
& wFLM &  & 0.796 (0.084) & 0.636 (0.042) & 0.614 (0.039)\\
\midrule
\multirow{8}{*}{II} & FLM & \multirow{2}{*}{50} & 0.084 (0.029) & 0.078 (0.027) & 0.076 (0.025)\\
& wFLM &  & 0.057 (0.026) & 0.044 (0.024) & 0.038 (0.021)\\
\cmidrule{2-6}
& FLM & \multirow{2}{*}{100} & 0.065 (0.018) & 0.059 (0.014) & 0.058 (0.014)\\
& wFLM &  & 0.041 (0.019) & 0.024 (0.014) & 0.022 (0.013)\\
\cmidrule{2-6}
& FLM & \multirow{2}{*}{200} & 0.052 (0.011) & 0.049 (0.008) & 0.049 (0.008)\\
& wFLM &  & 0.031 (0.016) & 0.015 (0.008) & 0.013 (0.007)\\
\cmidrule{2-6}
& FLM & \multirow{2}{*}{500} & 0.045 (0.004) & 0.044 (0.004) & 0.043 (0.004)\\
& wFLM &  & 0.022 (0.013) & 0.010 (0.003) & 0.008 (0.003)\\
\bottomrule
\end{tabular}
\caption{Average mean squared prediction errors and standard deviations (in parentheses) for bounded domain $\mathcal{T}=[0,1]$, based on 1000 Monte Carlo simulations, comparing FLM (functional linear model with Lebesgue measure) and wFLM (weighted functional linear model with step function weights).}
\label{tab:bounded}
\end{table}



\subsection{Simulations on $\mathcal{T}=[0,\infty)$}\label{subsec:unbounded}

To evaluate the performance of the proposed weighted functional linear model (wFLM) on an unbounded domain, we conducted a simulation study over $\mathcal{T} = [0,\infty)$. In addition to the exponential density, which aligns with the true underlying measure used to generate the data, we also included the half-normal density to assess the robustness of the method regarding the choice of weighting function. We compared the predictive performance of wFLM with exponential weight function, wFLM with half-normal weight function and the classical FLM that utilizes the Lebesgue measure. We considered four settings with sample sizes $n$ ranging from $\{100, 200\}$ and two choices for the number of measurements $N_i$ with $Q=500$ Monte Carlo runs. The number of measurements $N_i$ was either set to $20$ or chosen randomly for each predictor trajectory with equal probability from $\{5,6,7,8,9,10\}$. The locations of the measurements were exponentially distributed with a rate $1/2$ over the infinite interval $[0,\infty)$, reflecting unbounded support with irregular and potentially sparse sampling. 



The predictor trajectories $X_i(\cdot)$ and associated noisy measurements $X_{i j}$ were generated as follows. The simulated processes $X$ had  mean function $\mu_X(s)=5 - 3\cos(\pi t/5) + 2t$ and covariance function constructed using a set of 9 eigenfunctions $\psi_{k}$ (for more details we refer to  Section S.4 in the Supplementary Material), which are orthonormal  in $L^2([0,\infty),d\nu)$ where $d\nu(t) = e^{-t} dt$ is the density of the standard exponential distribution. We chose the eigenvalues  $\rho_{wk}=10\times 0.5^{k-1}, \,\, k=1,\ldots,9$ and $\rho_{wk}=0, k > 9$, and $\sigma_X^2 = 0.5$ as  variance of the additional measurement errors $\epsilon_{i j}$, which were assumed to be normal with mean 0. For each sample $i$, we generate FPC scores $\xi_{ik}$ from $N(0,\rho_k)$ and obtained   predictor measurements, $X_{ij}=\mu_X(t_{ij}) + \sum_{k=1}^9\xi_{ik}\psi_{k}(t_{ij})+\epsilon_{ij},\,\, j=1,\ldots,N_i, \,\, i=1,\ldots,n$. The scalar responses $Y_i$ were generated by $Y_i = \sum_{k=1}^9 \beta_k\xi_{ik} + e_i$, where $\beta_k = 10/k^3$ and $e_i\stackrel{i.i.d.}{\sim}N(0, 0.5^2)$. As before, for each Monte Carlo run we generated 100 new noisy predictors $X_{ij}^*$ and 100 corresponding noise-free responses $Y_i^*$.

Table~\ref{tab:exp} reports the average mean squared prediction errors and standard deviations across the simulations. Here both wFLM approaches dramatically outperform the classical FLM for the irregular and sparse measurement settings. For instance, when $n=100$ and $N_i \in {5,\ldots,10}$, wFLM (Exp) reduces AMSPE by 88.4\% compared to the basic FLM, while wFLM (Half-Normal) yields similar gains (89.0\%), despite the fact that the true data-generating measure here is the standard exponential density. This demonstrates a certain robustness of our method regarding  the specific choice of a weight function derived from a parametric distribution.  Even in relatively irregular dense settings, e.g., when $N_i = 20$,  both weighting schemes substantially improve prediction accuracy. As expected, performance improves with larger sample size $n = 200$ and both wFLM approaches maintain a clear advantage over FLM. These results highlight the flexibility and reliability of the proposed framework for unbounded domains and irregular measurement patterns. 

\begin{table}[tb]
\centering
\begin{tabular}{lccc}
\toprule
Method & $n$ & $N_i$= 5-10 & $N_i$ = 20\\
\midrule
FLM & \multirow{3}{*}{100} & 1000.49 (1267.99) & 782.28 (1063.72)\\
wFLM (Exp) &  & 116.09 (388.10) & 61.53 (435.21)\\
wFLM (HalfNorm) &  & 105.80 (478.56) & 52.83 (385.37)\\
\midrule
FLM & \multirow{3}{*}{200} & 804.27 (1016.61) & 678.97 (956.70)\\
wFLM (Exp)&  & 82.08 (240.86) & 31.27 (330.86)\\
wFLM (HalfNorm) &  & 76.13 (187.65) & 39.82 (324.44)\\
\bottomrule
\end{tabular}
\caption{Average mean squared prediction errors and standard deviations (in parentheses) for unbounded domain $\mT=[0,\infty)$, based on 500 Monte Carlo simulations, comparing the FLM (functional linear model with Lebesgue measure) and wFLM (weighted functional linear model with exponential density weights and half-normal density weights).}
\label{tab:exp}
\end{table}

\section{Applications}
\subsection{Predicting COVID-19 new cases}\label{app:covid}

We illustrate the performance of the proposed method with COVID-19 data. Functional data analyses for time-dynamic data of COVID-19 cases have been conducted previously \citep{carr:20, mull:22:2}. We obtained daily confirmed cases across countries from the COVID-19 Data Repository by the Center for Systems Science and Engineering (CSSE) at Johns Hopkins University. These data are publicly available at \url{https://github.com/CSSEGISandData/COVID-19}. The data feature the cumulative number of confirmed cases for each country from January 22, 2020, to March 3, 2023 and were accessed on April 11, 2023. For the analysis, we focused on the period from July 1, 2020, to December 31, 2022 (a total of 914 days), and used the seven-day moving average of daily confirmed cases per million as functional predictor. The scalar response was taken as the total confirmed cases from January 1 to January 31 in 2023. 

\begin{figure}[tb]
    \centering
    \includegraphics[width=.65\linewidth]{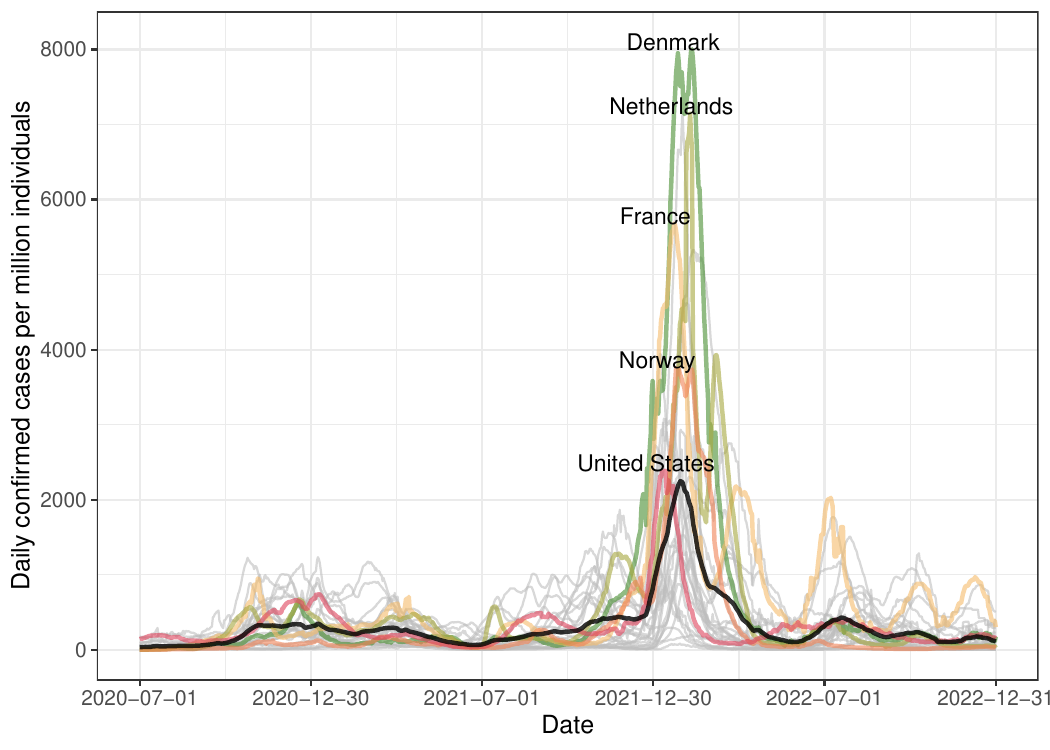}
    \caption{Trajectories of the seven-day moving average of daily confirmed cases per million individuals for 29 countries. Smoothed mean curves are marked by the bold black line.}
	\label{fig: daily confirmed cases for 29 countries}
   \end{figure}

The seven-day moving averages of daily confirmed cases per million people from July 1, 2020 to December 31, 2022 for 29 countries are displayed in Fig \ref{fig: daily confirmed cases for 29 countries}. The 29 countries for which data were included were located in America or Europe, as they exhibited similar COVID-19 response policies and relatively low bias in reported cases, and countries with zero cases in the 914-day trajectory were excluded. 
We investigated the performance of both the selection of a weight function as a step function and as an exponential density. We applied the infinite domain method here as the functional predictor spans a fairly long duration and as it is reasonable to assume that the closer the data points are to the end of 2022, the more crucial their influence becomes for the subsequent total confirmed cases in January 2023. Therefore, it is reasonable to assign higher importance to the domain at the end of 2022 while assigning relatively lower weight to the data for earlier periods. To implement this strategy within the framework of a weight function derived from the exponential distribution, we encoded December 31, 2022, as $t = 1$ and July 1, 2020, as $t = 914$. This coding scheme ensured that data with measurement times closer to $t = 1$ received more weight than those measured earlier.  
In contrast, when implementing weight functions as  step functions we retained the original domain, with $t = 1$ representing July 1, 2020 and $t = 914$ representing December 31, 2022.

To compare the performance of the FLM with weight function selection with the original FLM that uses the Lebesgue measure and therefore a constant weight function,  we employed a leave-one-out cross-validation score, see Table \ref{tab:covid}. The leave-one-out cross-validation score is 
$$\text{LOOCVS} = \frac{\sum_{i=1}^{n}(Y_i - \hat{Y}_i^{(-i)})^2}{\sum_{i=1}^n(Y_i-\bar{Y})^2},$$ 
where $\hat{Y}_i^{(-i)}$ represents the predicted value from the model after omitting the $i$th country from the training data. 

The optimal number of principal components, i.e., the minimizer of the cross-validation score  for the classical FLM was found to be $\text{M}_{Leb} = 4$, while the optimal M for the step weight function and the weight function derived from the exponential distribution were  3 and 2. Table \ref{tab:covid} reveals that wFLM (step) and wFLM (exp) achieve better prediction performance, resulting in a $34\%$ and $66\%$ improvement in prediction accuracy compared to the classical FLM  with the Lebesgue measure.  It clearly emerges that wFLM achieves better prediction accuracy in this application while using fewer principal components as compared to FLM. 

\begin{table}
\caption{Leave-one-out cross-validation score and standard deviations (in parentheses) of wFLM (weighted functional linear model with step function weight and exponential density weight function) and FLM (functional linear model with Lebesgue measure) for COVID-19 data}
\label{tab:covid}
\centering
	\begin{tabular}{ccc}
\toprule
wFLM (Step) & wFLM (Exp) & FLM (Lebesgue)\\
\midrule
0.513 (0.809) & 0.263 (0.328) & 0.779 (1.027)\\
\bottomrule
\end{tabular}
\end{table}

\subsection{National Health and Nutrition Examination Survey}
\label{app:NHANES}

Behavioral scientists are interested in analyzing the association between cardiovascular risk factors (such as systolic blood pressure and total cholesterol) and physical activity \citep{luke:11, gera:15, ledb:22, ge:24}. We apply the proposed method to model the effect of physical activity intensity on systolic blood pressure, utilizing data from the National Health and Nutrition Examination Survey (NHANES) 2005-2006; these data are publicly available at  \url{https://wwwn.cdc.gov/nchs/nhanes/ContinuousNhanes/Default.aspx?BeginYear=2005}. NHANES assesses the health and nutrition status of U.S. adults and children through comprehensive interviews and physical examinations. The survey collects information on demographic, socioeconomic, dietary, and health-related variables, along with medical, dental and physiological measurements. As part of NHANES, participants aged six and older were asked to wear an Actigraph 7164 accelerometer on a waist belt for seven consecutive days, capturing physical activity intensity every minute throughout the day. These accelerometer data have been widely used by researchers to explore the relationship between activity patterns and various health indicators \citep{troi:14, tudo:12}. Worn on the right hip, the accelerometer began recording at 12:01 am the day after the participant’s health examination and was removed only during sleep, swimming or bathing.

We restricted our analysis to a subset of male participants who were married, aged over 20 and had four complete blood pressure measurements. This led to a sample size of $n=500$ participants. Denoting the physical activity intensity function at minute $t$ of the $i$th participant by  $U_i(t)$, we observe that its domain is $I_i=\{t\in[0, 10080]: 1\leq U_i(t)\}$, where $10080$ is in minutes and stands for the total number of minutes over the $7$ days where the signal was recorded.  We transform the $U_i$  to define the  predictor for the $i$th subject as $X_i(s) = \#\{t\in I_i: U_i(t) = s\}$ for a given physical activity intensity level $s > 0$. This function represents the total time in minutes during which the physical activity intensity equals $s$ over the $7$ days of observation. This is possible since the activity levels are discrete.  Similar transformations of the physical activity intensity have been considered previously by various authors   \citet{chan:20:2, lin:23}, as this transformed function provides a better reflection of the actual activity than $U_i(t)$ does.   

The response of interest is the average systolic blood pressure, averaging over the four available measurements. The potential for large values of physical activity intensity $s$, which serves as argument of the $X_i$, means that the domain has no clear upper bound, motivating to consider an infinite domain $(0,\infty)$ for the functional predictor. We investigate the performance of the proposed wFLM with a weight function derived from the exponential standard distribution in comparison with the ordinary FLM.  Here it is reasonable to implement the exponential density weight, as the majority of the physical intensity values are small. It is natural that most of the time, people will engage in sedentary behavior or light physical activity, and rarely have high-intensity values and therefore low levels of physical activity should receive more weight. By cross-validation we found the optimal number of principal components for the ordinary FLM to be $\text{M}_{Leb} = 1$. With $\text{M} = 1$, Table \ref{tab:nhanes} reveals that wFLM (exp) achieves better prediction performance, resulting in a $27\%$ improvement in prediction accuracy compared to FLM (Lebesgue).

\begin{table}
\caption{Leave-one-out cross-validation score and standard deviations (in parentheses) of wFLM (weighted functional linear model with exponential density weight function) and FLM (functional linear model with Lebesgue measure) for NHANES data.}
\label{tab:nhanes}
\centering
	\begin{tabular}{cc}
\toprule
wFLM (Exp) & FLM (Lebesgue)\\
\midrule
1.022 (1.633) & 1.398 (8.814)\\
\bottomrule
\end{tabular}
\end{table}

\section{Discussion}
In this paper, we introduced a weighted functional linear model that generalizes the conventional functional linear model by incorporating a data-driven, optimal measure for defining the Hilbert space. This modified model is shown to achieve better predictive performance by emphasizing more relevant regions of the functional domain and thus improving the representation of the coefficient function. Furthermore, this framework naturally extends to infinite domains, addressing challenges in functional data analysis for unbounded domains where traditional methods may struggle. 

Through simulation studies and real data applications, we demonstrated that the approach consistently outperforms the standard functional linear model, offering a more flexible and powerful framework for functional linear regression.  We also provide basic representations and relationships of eigen-systems for weighted and unweighted functional principal component analysis. 

The proposed method could also be harnessed for other tasks in functional data analysis,  such as generalized functional linear regression and functional classification \citep{mull:05:1}, among others.   Future work may explore alternative methods for selecting optimal measures and expanding the model to accommodate more complex functional data structures.

\section*{Acknowledgments}
\label{sec:akd}
This research was supported in part by NSF grant DMS-2310450. We thank the referees for helpful comments. 

\bibliographystyle{elsarticle-harv} 
\bibliography{collection.bib}

\newpage
\bigskip
\begin{center}
{\large\bf Supplementary Material}
\end{center}
\setcounter{subsection}{0}
\setcounter{table}{0}
\renewcommand{\thesubsection}{S.\arabic{subsection}}
\renewcommand{\thetable}{S.\arabic{table}}
\setcounter{prop}{0}
\setcounter{lem}{0}
\renewcommand{\theprop}{S\arabic{prop}}
\renewcommand{\thelem}{S\arabic{lem}}

\subsection{Proof of Proposition \ref{prop:eigen-system-Z}}
\begin{proof}
We show that $\phi_{wj}(t)=\phi_{Zj}(t)/\sqrt{w(t)}$ form an orthonormal system with respect to $d \nu$,
	\begin{align*}
    \int_{\mathcal{T}}\phi_{wk}^2(t) d\nu(t) &= \int_\mathcal{T}\frac{\phi^2_{Zk}(t)}{w(t)}w(t) dt = 1,\\
    \int_{\mathcal{T}} \phi_{wj}(t)\phi_{wk}(t) d\nu(t) &= \int_{\mathcal{T}} \frac{\phi_{Zj}(t)}{\sqrt{w(t)}}\frac{\phi_{Zk}(t)}{\sqrt{w(t)}}w(t)dt = 0, \text{ for } j\ne k.
\end{align*}
Denoting the auto-covariance operator of $X$ with respect to $d \nu$, i.e., in the space $L^2(\mathcal{T}, \nu)$ by $A_w$.
$$
\left\{A_w(f)\right\}(t)=\int_{\mathcal{T}} C_{XX}(s, t) f(s) d\nu(s).
$$
Next we  show that the $\phi_{w j}$ and $\rho_{wj}$ are the eigenfunctions and eigenvalues of $A_w$,
Then
\begin{equation*}
    \begin{split}
        \{A_w(\phi_{w k} )\}(t) & = \int_{\mT} C_{XX}(s,t) \phi_{w k}(s) d \nu(s) \\
        & = \int_{\mT} \frac{1}{\sqrt{w(t) w(s)}} \cov\{Z(s), Z(t)\}\frac{\phi_{Z k}(s)}{\sqrt{w(s)}} w(s) d s \\
        & =\frac{1}{\sqrt{w(t)}} \sum_{j = 1}^\infty \rho_{wj} \phi_{Z j}(t) \int_{\mT} \phi_{Z k}(s) \phi_{Z j}(s) d s \\
        & =\rho_{wk} \frac{\phi_{Z k}(t)}{\sqrt{w(t)}} = \rho_{wk} \phi_{w k}(t).
    \end{split}
\end{equation*}
Finally, it is easy to show that the functional principal component scores of process $X$ in $L^2(\mathcal{T}, \nu)$ are the same as those of process $Z$ in $L^2(\mathcal{T})$,
$$\xi_{wk} = \int_{\mT} Z(t)\phi_{Zk}(t) dt = \int_{\mT}\{X(t)-\mu_X(t)\}\phi_{wk}(t)d\nu(t) .$$  
\end{proof}

\subsection{Proof of Proposition \ref{prop:eigensystem}}
\begin{proof}
Let $w(t)$ be $d\nu_2(t)/d\nu_1(t)$. We first show that $\psi_{wk}(t)=\frac{\psi_{Zk}(t)}{\sqrt{w}}$ form an orthonormal system with respect to $d \nu_2$,
	\begin{align*}
    \int_{\mT}\psi_{wk}^2(t) d\nu_2(t) &= \int_\mT\frac{\psi^2_{Zk}(t)}{w(t)} w(t) d\nu_1(t) = 1,\\
    \int_{\mT} \psi_{wj}(t)\psi_{wk}(t) d\nu_2(t) &= \int_{\mT} \frac{\psi_{Zj}(t)}{\sqrt{w(t)}}\frac{\psi_{Zk}(t)}{\sqrt{w(t)}}w(t)d\nu_1(t) = 0, \text{ for } j\ne k,
\end{align*}
since $\{\psi_{Zk}\}_{k=1}^\infty$ are eigenfunctions of the process $Z = \sqrt{w}(X - \mu_X)$. Denoting the auto-covariance operator of $X$ with respect to $d \nu_2$, i.e., in the space $L^2(\mathcal{T}, \nu_2)$ by $A_w$.
$$
\left\{A_w(f)\right\}(t)=\int_{\mathcal{T}} C_{XX}(s, t) f(s) d\nu_2(s).
$$
Next we show that the $\psi_{w k}$ and $\rho_{wk}$ are the eigenfunctions and eigenvalues of $A_w$.
For this we observe 
\begin{equation*}
    \begin{split}
        \{A_w(\psi_{w k} )\}(t) & = \int_{\mT} C_{XX}(s,t) \psi_{w k}(s) d \nu_2(s) \\
        & = \int_{\mT} \frac{1}{\sqrt{w(t) w(s)}} \cov(Z(s), Z(t))\frac{\psi_{Z k}(s)}{\sqrt{w(s)}} w(s) d \nu_1(s) \\
        & =\frac{1}{\sqrt{w(t)}} \sum_{j = 1}^\infty \rho_{wj} \phi_{Z j}(t) \int_{\mT} \psi_{Z k}(s) \psi_{Z j}(s) d \nu_1(s) \\
        & =\rho_{wk} \frac{\psi_{Z k}(t)}{\sqrt{w(t)}} = \rho_{wj} \psi_{w k}(t).
    \end{split}
\end{equation*}
Finally, it is easy to show that the functional principal component scores of process $X$ in $L^2(\mathcal{T}, \nu_2)$ are the same as those of process $Z$ in $L^2(\mathcal{T}, \nu_1)$,
$$\xi_{wk} = \int_{\mathcal{T}} Z(t)\phi_{Zk}(t) d\nu_1 = \int_{\mT}\{X(t)-\mu_X(t)\}\psi_{wk}(t)d\nu_2(t) .$$  
\end{proof}

\subsection{Dyadic splitting algorithm\label{app:dyadic}}

\textbf{Initialization:} In the first step, we divide the interval $[0, 1]$ into two subintervals: $I_{{1}} = [0,1/2)$ and $I_{{2}}=[1/2,1]$. We seek a constant weight $w_{{1}}$ for $I_{{1}}$ that minimizes the cross-validation mean square prediction error. The weight $w_{{2}}$ for $I_{{2}}$ is determined automatically based on the constraints imposed on the weight function.

\textbf{Refinement:} Following the initialization, we possess weights for both $I_{{1}}$ and $I_{{2}}$. We further split $I_{{1}}$ into two equal subintervals: $I_{{1,1}}=[0,1/4)$ and $I_{{1,2}}=[1/4,1/2)$. While keeping $w_{{2}}$ unchanged on $I_{{2}}$, we search for a weight $w_{{1,1}}$ on $I_{{1,1}}$ as in the first step, with $w_{{1,2}}$ automatically determined.\\
We then perform a similar procedure for $I_{{2}}$ from the initialization step, splitting it into $I_{{2,1}}=[1/2,3/4)$ and $I_{{2,2}}=[3/4,1]$, while retaining the weights on the other intervals. This results in weights $w_{{2,1}}$ on $I_{{2,1}}$ and $w_{{2,2}}$ on $I_{{2,2}}$, automatically adjusted based on the constraints.

\textbf{Updating Step:} At the $k$th step, where there are $2^{k-1}$ intervals, we iteratively split each interval from the previous step at its midpoint. We determine the corresponding weights as constants on the left and right subintervals, aiming to minimize the cross-validation mean square prediction error.

\textbf{Termination:} The iteration continues until further splitting fails to reduce the cross-validation mean square prediction error or until it reaches the maximum allowable number of splitting steps (typically set to 3). At this point, we conclude the algorithm, and the current weight function is designated as the output.\\

\subsection{Othonormal basis function in $L^2([0,\infty),\nu)$ with $d\nu(t)=\lambda e^{-\lambda t} dt$\label{app:basis}}
\begin{align*}
	& \phi_1(t) = 1,\\
	& \phi_2(t) = 1-\lambda t,\\
	& \phi_3(t) = 1 -2\lambda t + \frac{\lambda^2}{2}t^2,\\
	& \phi_4(t) = 1 -3\lambda t + \frac{3\lambda^2}{2}t^2 - \frac{\lambda^3}{6}t^3,\\
	& \phi_5(t) = 1 -4\lambda t + \frac{6\lambda^2}{2}t^2 - \frac{4\lambda^3}{6}t^3 + \frac{\lambda^4}{24}t^4,\\
	& \phi_6(t) = 1 -5\lambda t + \frac{10\lambda^2}{2}t^2 - \frac{10\lambda^3}{6}t^3 + \frac{5\lambda^4}{24}t^4 - \frac{\lambda^5}{120}t^5,\\
	& \phi_7(t) = 1 -6\lambda t + \frac{15}{2}\lambda^2t^2 - \frac{20\lambda^3}{6}t^3 + \frac{5\lambda^4}{8}t^4 - \frac{\lambda^5}{20}t^5 + \frac{\lambda^6}{720}t^6,\\
	& \phi_8(t) = 1 -7\lambda t + \frac{21}{2}\lambda^2t^2 - \frac{35\lambda^3}{6}t^3 + \frac{35\lambda^4}{24}t^4 - \frac{7\lambda^5}{40}t^5 + \frac{7\lambda^6}{720}t^6 - \frac{\lambda^7}{5040}t^7,\\
	& \phi_9(t) = 1 -8\lambda t + \frac{28}{2}\lambda^2t^2 - \frac{56\lambda^3}{6}t^3 + \frac{70\lambda^4}{24}t^4 - \frac{7\lambda^5}{15}t^5 + \frac{7\lambda^6}{180}t^6 - \frac{\lambda^7}{630}t^7 + \frac{\lambda^8}{40320}t^8.
\end{align*}

\subsection{Training time and computational complexity}\label{app:runtime}

\begin{table}[tb]
\centering
\caption{Average training time in minutes across different sample sizes and different numbers of measurement points and standard deviations (in parentheses) for FLM (functional linear model with Lebesgue measure) and wFLM (weighted functional linear model with step function weights) under Scenario 1 of Section \ref{subsec:bounded}.}
\label{tab:runtime_step}
\begin{tabular}{ccccc}
\toprule
Method & $n$ &  $N = 20$ & $N = 50$ & $N = 100$ \\
\midrule
FLM & \multirow{2}{*}{50} & 0.032 (0.006) & 0.030 (0.004) & 0.033 (0.005)\\
wFLM &  & 0.180 (0.022) & 0.236 (0.017) & 0.422 (0.055)\\
\midrule
FLM & \multirow{2}{*}{100} & 0.030 (0.004) & 0.030 (0.003) & 0.035 (0.006)\\
wFLM &  & 0.224 (0.024) & 0.323 (0.024) & 0.693 (0.065)\\
\midrule
FLM & \multirow{2}{*}{200} & 0.031 (0.008) & 0.031 (0.004) & 0.040 (0.004)\\
wFLM &  & 0.342 (0.056) & 0.504 (0.051) & 1.223 (0.108)\\
\midrule
FLM & \multirow{2}{*}{500} & 0.033 (0.005) & 0.037 (0.011) & 0.057 (0.010)\\
wFLM &  & 0.667 (0.071) & 1.106 (0.246) & 2.785 (0.213)\\
\bottomrule
\end{tabular}
\end{table}

\begin{table}[tb]
\centering
\caption{Average training time in minutes across different sample sizes and different numbers of measurement points and standard deviations (in parentheses) for FLM (functional linear model with Lebesgue measure) and wFLM (weighted functional linear model with exponential density weight function) under the setting of Section 4.2.}
\label{tab:runtime_exp}
\begin{tabular}{cccc}
\toprule
Method & $n$ & $N_i = 5-10$ & $N_i = 20$ \\
\midrule
FLM & \multirow{2}{*}{100} & 0.076 (0.009) & 0.291 (0.041)\\
wFLM &  & 0.795 (0.125) & 4.395 (0.133)\\
\midrule
FLM & \multirow{2}{*}{200} & 0.218 (0.041) & 1.298 (0.147)\\
wFLM &  & 3.208 (0.126) & 17.759 (0.438)\\
\bottomrule
\end{tabular}
\end{table}

We present a detailed comparison of the training time for FLM and wFLM under various sample sizes and numbers of measurement points. Tables~\ref{tab:runtime_step} and~\ref{tab:runtime_exp} report average runtime (in minutes) over 100 repetitions, for wFLM (step) under a bounded domain and wFLM (Exp) under a unbounded domain, respectively. All computations were performed on a local machine equipped with an Apple M2 processor running macOS Sequoia.

Table~\ref{tab:runtime_step} reports the average training time for classical FLM and the step-based wFLM under Scenario 1 of Section~4.1, where the domain is bounded, $\mathcal{T} = [0,1]$. The results demonstrate that the step-based wFLM method introduces additional computational cost compared to FLM, but the increase is moderate and scales reasonably with both the sample size $n$ and the number of measurement points $N$. Table~\ref{tab:runtime_exp} presents training times for FLM and wFLM with exponential weights in the unbounded domain setting, $\mathcal{T} = [0, \infty)$, described in Section~4.2. Here, the computational cost is higher. There are two main reasons for this. First, the predictor measurements are irregularly spaced over an infinite interval, which increases computational burden compared to evenly spaced and bounded designs. Second, wFLM (Exp) involves grid searching over a set of candidate parameters, which adds further cost.

Although wFLM requires more computational resources, it consistently outperforms classical FLM in all simulation settings and applications. In particular, for unbounded domains, FLM suffers from fundamental theoretical limitations. The space $L^2([0,\infty))$ excludes many commonly used functions, such as polynomials, and thus cannot adequately support standard FPCA or FLM procedures. As shown in Section~4.2 and Section~\ref{supp:sensitivity_exp} of the Supplementary Material, FLM performs poorly in these scenarios, whereas wFLM remains stable and accurate. Moreover, even in bounded domains where the true underlying measure is the Lebesgue measure, as in Scenario 1 of Section~4.1 and Supplementary Section~\ref{supp:sensitivity_step}, wFLM still yields superior prediction accuracy when the leading functional principal components fail to capture the signal structure effectively. The additional computational cost of wFLM is therefore justified by its substantial gains in predictive performance and theoretical soundness.

\subsection{Sensitivity analysis regarding  measurement error variance for step-based wFLM}
\label{supp:sensitivity_step}

\begin{table}[tb]
\centering
\caption{Average mean squared prediction error (AMSPE) and standard deviations (in parentheses) for FLM (functional linear model with Lebesgue measure) across different sample sizes $n$, number of measurements $N$, and measurement error levels $\sigma$.}
\label{tab:amspe_sensitivity_flm}
\begin{tabular}{ccccc}
\toprule
$\sigma$ & $n$ & $N = 20$ & $N = 50$ & $N = 100$ \\
\midrule
\multirow{4}{*}{0.00} 
& 50  & 1.077 (0.168) & 1.077 (0.167) & 1.077 (0.166) \\
& 100 & 0.998 (0.140) & 0.999 (0.140) & 0.999 (0.140) \\
& 200 & 0.983 (0.159) & 0.983 (0.159) & 0.983 (0.158) \\
& 500 & 0.968 (0.135) & 0.968 (0.135) & 0.968 (0.135) \\
\midrule
\multirow{4}{*}{0.25} 
& 50  & 1.086 (0.171) & 1.083 (0.178) & 1.090 (0.170) \\
& 100 & 1.009 (0.144) & 1.003 (0.143) & 1.008 (0.143) \\
& 200 & 0.983 (0.155) & 0.983 (0.158) & 0.988 (0.157) \\
& 500 & 0.971 (0.136) & 0.970 (0.136) & 0.971 (0.136) \\
\midrule
\multirow{4}{*}{0.50} 
& 50  & 1.086 (0.172) & 1.082 (0.177) & 1.089 (0.169) \\
& 100 & 1.010 (0.143) & 1.003 (0.143) & 1.008 (0.143) \\
& 200 & 0.983 (0.155) & 0.983 (0.158) & 0.988 (0.157) \\
& 500 & 0.971 (0.136) & 0.970 (0.136) & 0.971 (0.136) \\
\midrule
\multirow{4}{*}{0.75} 
& 50  & 1.088 (0.173) & 1.082 (0.179) & 1.088 (0.168) \\
& 100 & 1.010 (0.143) & 1.003 (0.143) & 1.009 (0.145) \\
& 200 & 0.982 (0.155) & 0.983 (0.158) & 0.988 (0.157) \\
& 500 & 0.971 (0.136) & 0.970 (0.135) & 0.971 (0.136) \\
\midrule
\multirow{4}{*}{1.00} 
& 50  & 1.087 (0.173) & 1.082 (0.177) & 1.087 (0.167) \\
& 100 & 1.010 (0.145) & 1.003 (0.142) & 1.009 (0.145) \\
& 200 & 0.983 (0.156) & 0.984 (0.158) & 0.987 (0.157) \\
& 500 & 0.972 (0.136) & 0.970 (0.135) & 0.971 (0.136) \\
\bottomrule
\end{tabular}
\end{table}

\begin{table}[tb]
\centering
\caption{Average mean squared prediction error (AMSPE) and standard deviations (in parentheses) for wFLM (weighted functional linear model with step function weights) across different sample sizes $n$, number of measurements $N$, and measurement error levels $\sigma$.}
\label{tab:amspe_sensitivity_wflm}
\begin{tabular}{ccccc}
\toprule
$\sigma$ & $n$ & $N = 20$ & $N = 50$ & $N = 100$ \\
\midrule
\multirow{4}{*}{0.00} 
& 50  & 0.986 (0.177) & 0.858 (0.265) & 0.784 (0.275) \\
& 100 & 0.863 (0.172) & 0.634 (0.206) & 0.605 (0.197) \\
& 200 & 0.825 (0.160) & 0.522 (0.126) & 0.511 (0.102) \\
& 500 & 0.806 (0.145) & 0.496 (0.076) & 0.496 (0.076) \\
\midrule
\multirow{4}{*}{0.25} 
& 50  & 0.987 (0.192) & 0.843 (0.247) & 0.814 (0.248) \\
& 100 & 0.881 (0.179) & 0.661 (0.182) & 0.637 (0.182) \\
& 200 & 0.836 (0.168) & 0.565 (0.092) & 0.564 (0.119) \\
& 500 & 0.812 (0.148) & 0.553 (0.084) & 0.543 (0.084) \\
\midrule
\multirow{4}{*}{0.50} 
& 50  & 0.993 (0.190) & 0.922 (0.208) & 0.938 (0.216) \\
& 100 & 0.896 (0.175) & 0.768 (0.170) & 0.746 (0.179) \\
& 200 & 0.840 (0.161) & 0.671 (0.103) & 0.651 (0.121) \\
& 500 & 0.823 (0.147) & 0.645 (0.095) & 0.623 (0.095) \\
\midrule
\multirow{4}{*}{0.75} 
& 50  & 1.009 (0.188) & 0.971 (0.193) & 0.977 (0.184) \\
& 100 & 0.901 (0.165) & 0.862 (0.166) & 0.843 (0.169) \\
& 200 & 0.855 (0.159) & 0.772 (0.131) & 0.749 (0.135) \\
& 500 & 0.837 (0.145) & 0.719 (0.107) & 0.696 (0.104) \\
\midrule
\multirow{4}{*}{1.00} 
& 50  & 1.017 (0.191) & 0.997 (0.188) & 0.995 (0.180) \\
& 100 & 0.917 (0.165) & 0.888 (0.160) & 0.874 (0.162) \\
& 200 & 0.867 (0.158) & 0.819 (0.127) & 0.806 (0.138) \\
& 500 & 0.853 (0.144) & 0.775 (0.114) & 0.760 (0.110) \\
\bottomrule
\end{tabular}
\end{table}

To evaluate the robustness of the proposed weighted functional linear model with step function weight under varying levels of measurement error, we conducted additional simulations based on Scenario 1 in Section 4.1. The functional predictors $X_i(t)$ were generated using the same mean and covariance structure as described in the main simulation setting. Specifically, the mean function was defined as
$$
\mu_X(t) = 2t - 5 \cos(2\pi t),
$$
and the covariance function was constructed using 10 eigenfunctions $\{\psi_k\}_{k=1}^{10}$ with corresponding eigenvalues $\rho_k = 10 \times 0.5^{10-k}$. The eigenfunctions were given by
$$
\psi_k(t) = 
\begin{cases}
\sqrt{2} \cos(k\pi t), & \text{for } k \in \{1,3,5,7,9\}, \\
\sqrt{2} \sin((k-1)\pi t), & \text{for } k \in \{2,4,6,8,10\}.
\end{cases}
$$
For each subject $i$, the functional trajectory was constructed as
$$
X_{ij} = \mu_X(t_{ij}) + \sum_{k=1}^{10} \xi_{ik} \psi_k(t_{ij}) + \epsilon_{ij},
$$
where $\xi_{ik} \sim \mathcal{N}(0, \rho_k)$ and $\epsilon_{ij} \sim \mathcal{N}(0, \sigma^2)$ represent i.i.d.  measurement errors. We examined five levels of measurement error variance: $\sigma \in \{0, 0.25, 0.5, 0.75, 1.0\}$. The scalar responses were generated via the functional linear model:
$$
Y_i = \sum_{k=1}^{10} \beta_k \xi_{ik} + e_i,
$$
where $\beta_k = 5 \times 0.5^{k-1}$ and $e_i \sim \mathcal{N}(0, 0.5^2)$.

We compared FLM and wFLM across sample sizes $n \in \{50, 100, 200, 500\}$ and grid resolutions $N \in \{20, 50, 100\}$. Prediction performance was measured using average mean squared prediction error (AMSPE), averaged over 200 Monte Carlo replicates.

The results in Tables~\ref{tab:amspe_sensitivity_flm} and \ref{tab:amspe_sensitivity_wflm} show that wFLM consistently achieves lower prediction error than FLM across all levels of measurement error, sample sizes, and numbers of measurements. While FLM exhibits relatively stable performance as measurement error increases, its overall accuracy remains limited. In contrast, wFLM demonstrates strong predictive performance in low-noise settings and retains its advantage even as noise levels grow. These findings highlight the robustness of the proposed weighting scheme and underscore the benefit of adapting an optimized measure rather than the default Lebesgue measure across many settings.

\subsection{Sensitivity analysis regarding  measurement error variance for weight functions derived from parametric distributions}
\label{supp:sensitivity_exp}

\begin{table}[tb]
\centering
\caption{Average mean squared prediction error (AMSPE) and standard deviations (in parentheses) for FLM (functional linear model with Lebesgue measure) and wFLM (weighted functional linear model with exponential density weight function) under varying noise levels $\sigma$, for different sample sizes $n$ and number of measurements $N_i$.}
\label{tab:amspe_exp_sensitivity}
\begin{tabular}{ccccc}
\toprule
$\sigma$ & $n$ & Method & $N_i = 5$--$10$ & $N_i = 20$ \\
\midrule
\multirow{4}{*}{0.00} 
& \multirow{2}{*}{100} & FLM  & 836.47 (1167.38) & 800.22 (1340.22) \\
&                      & wFLM & 42.70 (64.91)    & 4.07 (3.83)      \\
\cmidrule{2-5}
& \multirow{2}{*}{200} & FLM  & 814.38 (1384.70) & 651.86 (968.43)  \\
&                      & wFLM & 32.38 (23.11)    & 2.61 (2.10)      \\
\midrule
\multirow{4}{*}{0.25} 
& \multirow{2}{*}{100} & FLM  & 788.45 (958.31)  & 824.89 (1279.89) \\
&                      & wFLM & 96.52 (452.15)   & 18.41 (111.21)   \\
\cmidrule{2-5}
& \multirow{2}{*}{200} & FLM  & 898.71 (1457.29) & 671.07 (1042.29) \\
&                      & wFLM & 40.83 (20.09)    & 5.37 (8.50)      \\
\midrule
\multirow{4}{*}{0.50} 
& \multirow{2}{*}{100} & FLM  & 912.36 (1156.95) & 802.88 (1164.14) \\
&                      & wFLM & 81.98 (112.51)   & 56.91 (412.74)   \\
\cmidrule{2-5}
& \multirow{2}{*}{200} & FLM  & 898.25 (1160.19) & 669.86 (958.95)  \\
&                      & wFLM & 66.28 (65.12)    & 17.03 (73.01)    \\
\midrule
\multirow{4}{*}{0.75} 
& \multirow{2}{*}{100} & FLM  & 942.42 (1182.26) & 873.84 (1357.36) \\
&                      & wFLM & 118.25 (220.73)  & 57.63 (269.89)   \\
\cmidrule{2-5}
& \multirow{2}{*}{200} & FLM  & 931.48 (1203.69) & 675.96 (924.48)  \\
&                      & wFLM & 110.87 (292.86)  & 35.00 (153.27)   \\
\midrule
\multirow{4}{*}{1.00} 
& \multirow{2}{*}{100} & FLM  & 1065.91 (1327.04)& 860.24 (1311.72) \\
&                      & wFLM & 143.09 (289.21)  & 86.20 (422.79)   \\
\cmidrule{2-5}
& \multirow{2}{*}{200} & FLM  & 924.58 (1142.45) & 690.19 (934.90)  \\
&                      & wFLM & 150.31 (503.18)  & 54.13 (254.85)   \\
\bottomrule
\end{tabular}
\end{table}

\begin{table}[tb]
\centering
\caption{Median mean squared error for FLM (functional linear model with Lebesgue measure) and wFLM (weighted functional linear model with exponential density weight function) under varying measurement error levels $\sigma$, for different sample sizes $n$ and numbers of measurements $N_i$.}
\label{tab:median_mse_exp_sensitivity}
\begin{tabular}{ccccc}
\toprule
$\sigma$ & $n$ & Method & $N_i = 5$--$10$ & $N_i = 20$ \\
\midrule
\multirow{4}{*}{0.00}
& \multirow{2}{*}{100} & FLM  & 559.87 & 372.53 \\
&                      & wFLM &  32.63 &   2.49 \\
\cmidrule{2-5}
& \multirow{2}{*}{200} & FLM  & 478.78 & 393.68 \\
&                      & wFLM &  25.68 &   1.93 \\
\midrule
\multirow{4}{*}{0.25}
& \multirow{2}{*}{100} & FLM  & 546.97 & 421.66 \\
&                      & wFLM &  44.14 &   4.09 \\
\cmidrule{2-5}
& \multirow{2}{*}{200} & FLM  & 472.03 & 366.03 \\
&                      & wFLM &  36.48 &   3.16 \\
\midrule
\multirow{4}{*}{0.50}
& \multirow{2}{*}{100} & FLM  & 648.08 & 423.55 \\
&                      & wFLM &  56.65 &   7.87 \\
\cmidrule{2-5}
& \multirow{2}{*}{200} & FLM  & 489.00 & 382.13 \\
&                      & wFLM &  50.19 &   5.69 \\
\midrule
\multirow{4}{*}{0.75}
& \multirow{2}{*}{100} & FLM  & 697.27 & 441.59 \\
&                      & wFLM &  70.63 &  12.33 \\
\cmidrule{2-5}
& \multirow{2}{*}{200} & FLM  & 544.98 & 393.73 \\
&                      & wFLM &  63.05 &  10.10 \\
\midrule
\multirow{4}{*}{1.00}
& \multirow{2}{*}{100} & FLM  & 797.91 & 454.03 \\
&                      & wFLM &  86.47 &  18.54 \\
\cmidrule{2-5}
& \multirow{2}{*}{200} & FLM  & 566.82 & 402.10 \\
&                      & wFLM &  78.71 &  15.21 \\
\bottomrule
\end{tabular}
\end{table}

To further assess the robustness of the proposed wFLM (Exp) method on the unbounded domain $[0,\infty)$, we conducted additional simulations based on the setting in Section 4.2, now incorporating varying levels of measurement error. Specifically, we varied the standard deviation of the additive noise $\epsilon_{ij} \sim \mathcal{N}(0, \sigma^2)$ with $\sigma \in \{0, 0.25, 0.5, 0.75, 1.0\}$. All other components of the data-generating process, including the eigenbasis functions $\{\psi_k\}$, the exponential measurement locations and the scalar response model  are  the same as described in Section 4.2.

We evaluated performance under two settings for the number of measurement points per trajectory: either fixed at $N_i = 20$ or randomly sampled from $\{5,6,7,8,9,10\}$ with equal probability. The locations of the measurements were exponentially distributed with a rate $1/2$ over the infinite interval $[0,\infty)$. Average mean squared prediction error (AMSPE) was computed over $Q = 200$ Monte Carlo runs for each setting and method.


Table~\ref{tab:amspe_exp_sensitivity} reports the average mean squared prediction error (AMSPE) and associated standard deviations. Across all settings, wFLM (Exp) consistently outperforms classical FLM, often by a large margin. Notably, while the prediction performance of FLM remains relatively stable across different noise levels, it is uniformly worse than wFLM, especially in low-noise or denser sampling regimes. In contrast, wFLM exhibits strong gains when the signal is recoverable, and  degrades modestly under increasing noise. To assess the impact of potential outliers in AMSPE due to heavy noise or extreme trajectories, we also report the median mean squared prediction error in Table~\ref{tab:median_mse_exp_sensitivity}. The results for median errors confirm and strengthen the trends observed in the results for mean errors:  wFLM achieves drastically lower median errors than FLM, particularly in the $N_i = 20$ setting, where the signal is better captured. These results demonstrate that the exponential weighting scheme not only improves prediction but also enhances robustness to measurement noise and sparsity.

\subsection{Sensitivity analysis regarding the choice of tuning parameters \texorpdfstring{$\lambda_1$ and $\lambda_2$}{lambda1 and lambda2}}\label{supp:lambda_sensitivity}

\begin{table}[tb]
\centering
\caption{Sensitivity of AMSPE to varying $\lambda_1$ (with $\lambda_2 = 0$ fixed) for step-based wFLM with $n=500$.}
\label{tab:sens_lambda1}
\begin{tabular}{ccccc}
\toprule
Method & $\lambda_1$ & $N = 20$ & $N = 50$ & $N = 100$ \\
\midrule
FLM & --- & 0.971 (0.136) & 0.970 (0.136) & 0.971 (0.136) \\
\midrule
\multirow{12}{*}{wFLM}  & 0.0 & 0.830 (0.143) & 0.645 (0.094) & 0.622 (0.094) \\
 & 0.1 & 0.938 (0.136) & 0.781 (0.201) & 0.770 (0.199)\\
 & 0.2 & 0.949 (0.134) & 0.948 (0.134) & 0.949 (0.134) \\
 & 0.3 & 0.951 (0.133) & 0.949 (0.130) & 0.950 (0.134) \\
 & 0.4 & 0.953 (0.135) & 0.950 (0.131) & 0.950 (0.134) \\
 & 0.5 & 0.953 (0.135) & 0.950 (0.131) & 0.950 (0.134) \\
 & 1.0 & 0.953 (0.135) & 0.950 (0.131) & 0.950 (0.134) \\
 & 1.5 & 0.954 (0.134) & 0.950 (0.132) & 0.950 (0.134) \\
 & 2.0 & 0.954 (0.134) & 0.950 (0.132) & 0.950 (0.134) \\
 & 3.0 & 0.954 (0.134) & 0.950 (0.132) & 0.950 (0.134) \\
 & 4.0 & 0.954 (0.134) & 0.950 (0.132) & 0.950 (0.134) \\
 & 5.0 & 0.954 (0.134) & 0.950 (0.132) & 0.950 (0.134) \\
\bottomrule
\end{tabular}
\end{table}

\begin{table}[tb]
\centering
\caption{Sensitivity of AMSPE to varying $\lambda_2$ (with $\lambda_1 = 0$ fixed) for step-based wFLM with $n=500$.}
\label{tab:sens_lambda2}
\begin{tabular}{ccccc}
\toprule
Method & $\lambda_2$ & $N = 20$ & $N = 50$ & $N = 100$ \\
\midrule
FLM & --- & 0.971 (0.136) & 0.970 (0.136) & 0.971 (0.136) \\
\midrule
\multirow{12}{*}{wFLM} & 0.0 & 0.830 (0.143) & 0.645 (0.094) & 0.622 (0.094) \\
 & 0.1 & 0.823 (0.146) & 0.645 (0.095) & 0.621 (0.095)\\
 & 0.2 & 0.823 (0.148) & 0.645 (0.095) & 0.621 (0.095)\\
 & 0.3 & 0.823 (0.147) & 0.645 (0.095) & 0.621 (0.095)\\
 & 0.4 & 0.823 (0.147) & 0.645 (0.095) & 0.621 (0.095)\\
 & 0.5 & 0.823 (0.147) & 0.645 (0.095) & 0.621 (0.095) \\
 & 1.0 & 0.823 (0.147) & 0.645 (0.095) & 0.621 (0.095) \\
 & 1.5 & 0.819 (0.145) & 0.645 (0.095) & 0.621 (0.095)\\
 & 2.0 & 0.819 (0.145) & 0.645 (0.095) & 0.621 (0.095)\\
 & 3.0 & 0.819 (0.145) & 0.645 (0.095) & 0.621 (0.095)\\
 & 4.0 & 0.819 (0.145) & 0.645 (0.095) & 0.621 (0.095)\\
 & 5.0 & 0.819 (0.145) & 0.645 (0.095) & 0.621 (0.095)\\
\bottomrule
\end{tabular}
\end{table}

To evaluate the sensitivity of the step-based wFLM regarding  the choice of tuning parameters $\lambda_1$ and $\lambda_2$, we conducted additional simulations under Scenario 1 of Section 4.1 in the main text. We fixed the sample size at $n = 500$ and considered three settings for the number of measurements $N \in \{20, 50, 100\}$. The measurement locations were equally spaced over $[0, 1]$, and the predictor trajectories $X_i(t)$ were generated with the same mean and covariance structure as described in the main text.

The parameter $\lambda_1$ penalizes the total variation of the step function weights, encouraging smoother transitions between adjacent intervals. A higher value of $\lambda_1$ flattens the weight function and for large values forces it to move closer to a uniform weight function and thus towards  the classical FLM. The parameter $\lambda_2$ penalizes the number of non-zero subintervals in the weight function. This promotes sparsity by allowing parts of the domain to be entirely down-weighted, which can help isolate the most informative regions and improve prediction.

We conducted two separate experiments. In the first, we fixed $\lambda_2 = 0$ and varied $\lambda_1$ from $0$ to $5$. In the second, we fixed $\lambda_1 = 0$ and similarly varied $\lambda_2$. The results are summarized in Tables~\ref{tab:sens_lambda1} and \ref{tab:sens_lambda2}, showing the average mean squared prediction error (AMSPE) and standard deviations over $200$ Monte Carlo replications.

When $\lambda_2$ is fixed at zero, increasing $\lambda_1$ leads to a noticeable increase in prediction error between $\lambda_1 = 0$ and $\lambda_1 = 0.1$, after which the performance stabilizes. This pattern suggests that even a small amount of total variation penalization can quickly push the step-function weight toward uniformity, diminishing its ability to adapt to the underlying signal. In this simulation setting, the Lebesgue measure is known to be suboptimal because the coefficient function $\beta(\cdot)$ cannot be effectively represented by the leading principal components. As $\lambda_1$ increases, the weight function becomes less adaptive and more uniform, resembling the classical FLM. This rigidity prevents the model from exploiting beneficial flexibility in weight function selection,  which explains the degradation in predictive performance. Conversely, when $\lambda_1$ is fixed at zero, increasing $\lambda_2$ leads to small gains in prediction accuracy. While the true underlying measure for this simulation is the Lebesgue measure, the inability of the leading principal components to capture $\beta(\cdot)$ motivates alternative weighting. By allowing the model to concentrate weight on more relevant subregions, the step-based wFLM is able to adapt better  to the signal structure.

Importantly, for users whose primary goal is predictive accuracy, we recommend setting $\lambda_1 = 0$, which removes the variation penalty and reduces computational burden. This configuration allows the model to explore more flexible weight structures without constraints. However, if interpretability of the learned measure is also a concern, such as avoiding abrupt shifts between adjacent intervals, a small positive $\lambda_1$ can help smooth the estimated weights. Overall, these findings suggest that the model is reasonably robust to tuning choices within a practical range, and the penalization framework provides users with the flexibility to balance prediction performance and interpretability depending on their analytical goals.

\end{document}